\documentclass[aps,prl,superscriptaddress,twocolumn,showpacs]{revtex4}%
\usepackage{amsmath, amsthm, amssymb}
\usepackage{graphicx}
\usepackage{color}
\usepackage{natbib}
\usepackage{hyperref}
\usepackage{tikz}
\usepackage{epstopdf}
\usepackage{amsmath}
\usepackage{amsfonts}
\usepackage{amssymb}%
\providecommand{\U}[1]{\protect\rule{.1in}{.1in}}
\providecommand{\U}[1]{\protect\rule{.1in}{.1in}}
\hypersetup{
colorlinks=true,
}

\usetikzlibrary{calc}
\usetikzlibrary{backgrounds}
\usetikzlibrary{arrows}
\usetikzlibrary{shapes.arrows}
\usetikzlibrary{decorations.markings}

\begin{document}
\title{$P$-wave Superfluidity by Blockade Effects in a Rydberg-Dressed Fermi Gas}
\email{stevenxiongbo@gmail.com}
\author{Bo Xiong}
\affiliation{Physics Department, National Tsing Hua University, Hsinchu, Taiwan}
\author{H. H. Jen}
\affiliation{Physics Department, National Tsing Hua University, Hsinchu, Taiwan}
\author{Daw-Wei Wang}
\affiliation{Physics Department, National Tsing Hua University, Hsinchu, Taiwan}
\affiliation{Physics Division, National Center for Theoretical Sciences, Hsinchu, Taiwan}
\affiliation{Frontier Research Center on Fundamental and Applied Sciences of Matter,
National Tsing Hua University, Hsinchu, Taiwan}

\begin{abstract}
We systematically investigate the $p$-wave superfluidity of a Rydberg-dressed
Fermi gas, where the soft-core effective interaction is of finite radius
$R_{c}$ due to blockade effects. After solving the BCS gap equation and
comparing the free energy, we obtain the quantum phase diagram, which is
composed of three different phases: polar ($p_{z}$), axial ($p_{x}+ip_{y}$),
and axi-planar ($p_{x}+i\beta_{p}p_{y}$) phases. The tri-critical point
locates around $R_{c}k_{F}\sim1$, where $k_{F}$ is the Fermi wave vector. We
further derive the Ginzburg-Landau theory to explain the phase diagram, and
estimate the transition temperature to be about 0.1$E_{F}$ in the current
experimental regime of $^{6}$Li. Our work paves the way for future studies on
$p$-wave superfluids and related quantum phase transitions in ultracold atoms.

\end{abstract}

\maketitle


\begin{figure}[t]
\begin{center}
\includegraphics[width = 0.99 \columnwidth]{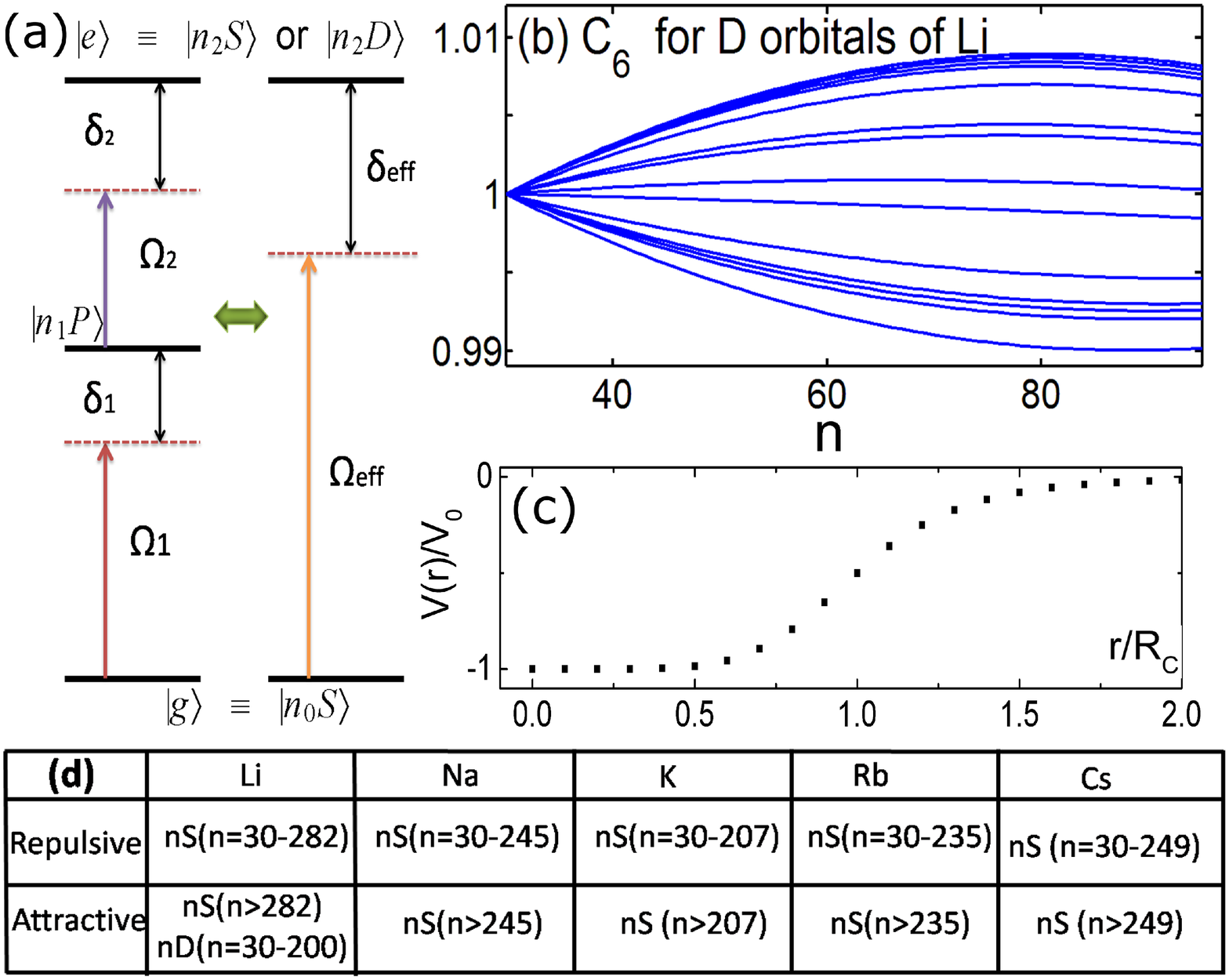}
\end{center}
\caption{(color online) (a) Schematic level plot for the off-resonant coupling
of the ground state ($\left\vert g\right\rangle $) to a Rydberg state
($\left\vert e\right\rangle $) via an intermediate state $|n_{1}P\rangle$. The
right panel is an effective single photon process with $\Omega_{\mathrm{eff}%
}=\Omega_{1}\Omega_{2}/2\delta_{1}$ and $\delta_{\mathrm{eff}}=\delta
_{1}+\delta_{2}$. (b) $nD-nD$ asymptote of Li atoms calculated by the quantum
defect theory \cite{SingerJPB2005,Walker08} and normalized by their value at
$n=30$. (c) The obtained effective interaction between Rydberg-dressed atoms,
where $V_{0}$ and $R_{c}$ are defined in the text. (d) Table of alkali atoms,
which may have an attractive or repulsive interaction by selecting a specific
quantum numbers in their Rydberg state \cite{SingerJPB2005,Walker08,QDT}.}%
\label{fig1}%
\end{figure}
{\textit{Introduction:}} Remarkable progress in the studies of ultracold
atomic Fermi gases \cite{Varenna,Trento} has opened up prospects for creating
novel phases not easily studied in traditional condensed matter systems. One
of the most important examples is the $p$-wave superfluid phase in identical
fermions \cite{Radzihovsky,Yip,Gurarie}, discussed in the contexts of
superfluid $^{3}$He \cite{Leggett,Volovik}, the fractional Quantum Hall effect
\cite{Green}, Sr$_{2}$RuO$_{4}$ \cite{MackenzieRMP2003}, and fermionic cold
atoms/molecules \cite{Shlyapnikov,Wang}. Besides of the non-trivial gap
symmetry, the topological $p$-wave pairing can be also degenerate in the
presence of vortices, spanned by zero-energy Majorana modes at vortex cores
\cite{Green,sternannals}. Such highly non-local character is expected to
suppress the decoherence, and may be used for topologically protected quantum
information processing \cite{sternannals}.

Among these candidates of $p$-wave superfluid phases, systems of ultracold
atoms are of special interest, because of the impurity-free environment and
the large flexibility for parameter detuning. However, this fascinating
$p$-wave superfluid phase usually has a very low critical temperature when
away from the Feshbach resonance, but a strong three-body loss when near the
resonance \cite{Jin}. Proposals utilizing the long-range nature of the dipolar
interaction between polar molecules are discussed recently
\cite{Shlyapnikov,Wang}, but the experimental realization of quantum
degenerate polar molecules is still challenging. On the other hand, recent
successful experiments in cold Rydberg gas with the blockade effects
\cite{SaffmanRMP2010,TongPRL2004,SingerPRL2004,RaithelPRL2005,SaffmanNP2009,GrangierNP2009,AdamsPRL2010,PfauPRL2007,PfauPRL2008A,PfauPRL2008B,PfauPRA2009,PfauPRL2012,KuzmichScience2012,LukinNature2012}
make it possible to realize a new strongly correlated system. Several
theoretical proposals
\cite{BuchlerPRL2008,ZollerPRL2010A,ZollerPRL2010B,PohlPRL2010,PohlPRL2012}
and experimental works \textbf{\cite{BlochNature2012}} have demonstrated the
possibility of generating a supersolid phase in bosonic Rydberg atoms with a
quiet long life time ($\sim0.5$ sec). Along this line, therefore, fermionic
Rydberg-dressed atoms (c.f. $^{6}$Li or $^{40}$K) should be a good candidate
for investigating unique properties not fully realized in other condensed
matter system, including the topological $p$-wave superfluid phase.

In this paper, we investigate the $p$-wave superfluid and its quantum phase
diagram of fermionic atoms in three dimensional (3D) space, where the
interatomic interaction is induced by weakly coupled to a Rydberg state
through a two-photon transition (see Fig. \ref{fig1}(a)). By self-consistently
solving the gap equation and comparing the total free energy, we can identify
the following three distinct phases: polar ($p_{z}$), axial ($p_{x}+ip_{y}$),
and axi-planar ($p_{x}+i\beta_{p}p_{y}$) states, where $\beta_{p}$ depends on
the relative momentum of the Cooper pair. The first two phases are
respectively equivalent to the $\beta$ and $A_{1}$ phases in $^{3}$He systems,
while the last one is a completely new phase. The tri-critical point is found
when the interaction range is about the inter-particle distance. The rich
quantum phase diagram can be qualitatively understood from Ginzburg-Landau
theory by considering the finite-range feature of the Rydberg-dressed
interaction, which can also enhance the transition temperature ($T_{c}$) to
the order of 0.1$E_{F}$ via nonlocal correlation effects. We then discuss the
parameter regime for observing these $p$-wave superfluid phases in the current
$^{6}$Li experiment.

{\textit{Interaction between Rydberg-dressed fermions:}} We consider identical
fermions in a 3D system (say $^{6}$Li or $^{40}$K). A far detuned two-photon
transition (see Fig. \ref{fig1}(a)) is applied to couple the ground state
($|n_{0}S\rangle$) to a Rydberg excited state ($|n_{2}S\rangle$ or
$|n_{2}D\rangle$) with an effective Rabi frequency ($\Omega_{\mathrm{eff}}$)
and detuning ($\delta_{\mathrm{eff}}$). Similar to the repulsive potential
considered in the bosonic case
\cite{Walker08,PohlPRL2011,PohlPRL2010,PfauPRL2010}, one may choose some
specific Rydberg states so that all the van der Waals scattering channels are
attractive ($C_{6}<0$), leading to the Cooper instability toward a superfluid
phase. Using results of quantum defect theory \cite{SingerJPB2005,Walker08},
we find that this condition can be fulfilled for $^{6}$Li when excited to
$|n_{2}D\rangle$ state ($n_{2}\geq$30) (see Fig. \ref{fig1}(b)) or to
$|n_{2}S\rangle$ state ($n_{2}\geq282$) \cite{QDT}. Available regime for both
pure attractive and repulsive interactions of other alkali atoms are also
shown together in Fig. \ref{fig1}(d).

Since alkali atoms usually have a relatively small fine structure splitting
(see Fig. \ref{fig1}(b)), the obtained effective interaction is almost
isotropic in 3D space \cite{PohlPRL2011,PohlPRL2010,PfauPRL2010}. The
effective interaction between Rydberg-dressed atoms can be calculated
perturbatively within the adiabatic approximation
\cite{Walker08,PohlPRL2011,PohlPRL2010,PfauPRL2010}:
\begin{equation}
V(\mathbf{r-r}^{\prime})=\frac{-V_{0}}{1+\left\vert \mathbf{r}-\mathbf{r}%
^{\prime}\right\vert ^{6}/R_{c}^{6}}, \label{DD1}%
\end{equation}
where $V_{0}\equiv\left(  \Omega_{\mathrm{eff}}/2\delta_{\mathrm{eff}}\right)
^{4}|\bar{C}_{6}|/R_{c}^{6}>0$. $R_{c}\equiv\left(  |\bar{C}_{6}%
|/2\delta_{\mathrm{eff}}\right)  ^{1/6}$ is the averaged soft-core radius
($\hbar\equiv1$), and $\bar{C}_{6}$ is some averaged van der Waals
coefficients after diagonalizing the whole scattering channels
\cite{PohlPRL2011}. Eq. (\ref{DD1}) is justified in the large detuning and
weak field limit, i.e. the number of effectively excited atoms
$N_{\mathrm{Ryd}}=(\Omega_{\mathrm{eff}}/2\delta_{\mathrm{eff}})^{2}N$ is
smaller than one, where $N$ is the total atom number.

We note that the effective interaction above has some unique and important
features for identical fermions we discussed here. (1) Such finite range
interaction strongly enhances the $p$-wave scattering amplitude, making it
possible to have a rapid evaporative cooling even without compensative cooling
with another species. (2) The adiabatic approximation used for the effective
potential, $V(\mathbf{r})$, should fail in the short-distance regime, because
of the complicated level crossing between different the electronic states.
However, such loss mechanism becomes less effective when considering identical
fermions due to the Pauli exclusion principle. (3) Finally, the weak
finite-range attractive interaction makes it possible to realize a stable
BCS-BEC quantum phase transition \cite{Radzihovsky,Yip} without a strong loss
rate in three-body collision, because the two-body bound state energy in the
BEC side can be so small if $V(\mathbf{r})$ is weak enough
\cite{Shlyapnikov,Wang}. As a result, we believe that identical fermions of
Rydberg-dressed interaction can be a better-controlled system for studying
anisotropic the $p$-wave superfluidity.

\begin{figure}[t]
\begin{center}
\includegraphics[width = 0.99 \columnwidth]{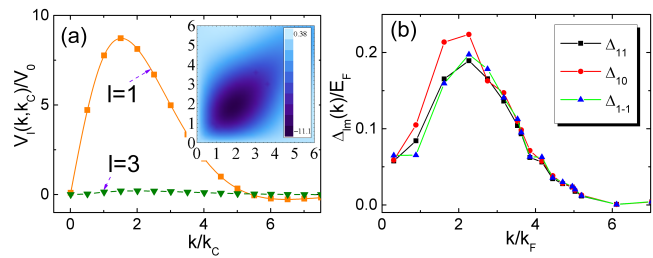}
\end{center}
\caption{(color online) (a) Interaction matrix element, $V_{l}(k,k_{c})$, with
$k_{c}\equiv R_{c}^{-1}$ for antisymmetric channels, $l=1$, $3$. The inset
shows the density plot of $V_{1}(k_{1},k_{2})$. (b) The calculated gap
function at zero temperature for different $p$-wave components. Note that the
maximum of the gap function is \textit{not} at Fermi surface, but determined
by the peak position of $V_{1}(k_{1},k_{2})$ as shown in (a). We choose
$\alpha=0.028$ and $R_{c}k_{F}=1.0$ as the point C in Fig. \ref{fig3}.}%
\label{fig2}%
\end{figure}

{\textit{Order parameter of $p$-wave superfluid:}} In this paper, we focus on
the superfluid phase in the BCS side without two-body bound state (see
Supplementary material for details). As a result, we can apply the first Born
approximation for the weak soft-core interaction, i.e. using $V_{\mathbf{k}%
-\mathbf{k}^{\prime}}=\int d\mathbf{r}V(\mathbf{r})\,e^{-i(\mathbf{k}%
-\mathbf{k}^{\prime})\cdot\mathbf{r}}$ as the scattering amplitude, so that
the BCS gap equation becomes ($\hbar=k_{B}\equiv1$):
\begin{equation}
\Delta_{\mathbf{k}}=-\frac{1}{\mathcal{V}}\sum_{\mathbf{k}^{\prime}}%
\frac{V_{\mathbf{k}-\mathbf{k}^{\prime}}}{2E_{\mathbf{k}^{\prime}}}%
\tanh\left(  \frac{E_{\mathbf{k}^{\prime}}}{2T}\right)  \Delta_{\mathbf{k}%
^{\prime}}, \label{reggap}%
\end{equation}
where $E_{\mathbf{k}}=\sqrt{\xi_{\mathbf{k}}^{2}+|\Delta_{\mathbf{k}}|^{2}}$
is the quasi-particle excitation spectrum with $\xi_{\mathbf{k}}\equiv
k^{2}/2M-\mu$ and $\Delta_{\mathbf{k}}=-\Delta_{-\mathbf{k}}$ being the
antisymmetric gap function. $M$ and $\mathcal{V}$ are the atom mass and system
volume. The chemical potential $\mu$ must be determined by the density
equation self-consistently: $n=\frac{1}{2\mathcal{V}}\sum_{\mathbf{k}}\left[
1-\frac{\xi_{\mathbf{k}}}{E_{\mathbf{k}}}\tanh\left(  \frac{E_{\mathbf{k}}%
}{2T}\right)  \right]  $.

Before solving above equations, we first decompose the interaction matrix
element and gap function into different angular momentum channels:
$V_{\mathbf{k}-\mathbf{k}^{\prime}}=-\sum_{lm}{V}_{l}\left(  k,k^{\prime
}\right)  Y_{l,m}(\hat{k})Y_{l,m}(\hat{k}^{\prime})$, and $\Delta_{\mathbf{k}%
}=\sum_{lm}^{\prime}\Delta_{l,m}(k)Y_{l,m}(\hat{k})$, where $\sum_{lm}%
^{\prime}$ is the summation over all odd values of $l$ and $m=-l,\cdots,l$. In
Fig. \ref{fig2}(a), we show the calculated $V_{l}(k,k^{\prime})$ and find that
the scattering amplitude is completely dominated by the $p$-wave channel. The
density plot of $V_{1}(k,k^{\prime})$ shows that it is peaked around
$k=k^{\prime}\simeq2R_{c}^{-1}$.

As for the gap function, it is argued that it can be well described by its
orientation alone \cite{Anderson61,Leggett} when considering pairing near the
Fermi surface in the weak interaction limit. Since the length scale of the
matrix element, $R_{c}$, is finite and may be comparable to the Fermi
momentum, here we need to consider a more general expression by separating the
"amplitude part" and the "orientation part" of the relative momentum:
\begin{equation}
\Delta_{\mathbf{k}}=\Delta(k)\left[  \sum_{m=0,\pm1}\beta_{k}^{m}Y_{1,m}%
(\hat{k})\right]  \equiv\Delta(k)\Psi_{k}(\hat{k}), \label{Delta_m}%
\end{equation}
where we keep the $p$-wave channel only and the complex coefficients,
$\{\beta_{k}^{m}\}$, is normalized as $\sum_{m}|\beta_{k}^{m}|^{2}=1$ (i.e.
$\int d\Omega|\Psi_{k}(\hat{k})|^{2}=1$) for all $k$. Since $p$-wave pairing
function generally break the gauge, time-reversal and rotational symmetries
when below $T_{c}$, we can classify them according to the residue symmetries
as discussed in the inert state of spinor condensate \cite{Yip_inert}: (1)
polar ($p_{z}$) state with $U(1)\times S^{2}$ symmetry (i.e. $\beta_{k}%
^{0}\neq0$ only), (2) axial ($p_{x}\pm ip_{y}$) state with $SO(3)$ symmetry
(i.e. $\beta_{k}^{\pm}\neq0$ only), and (3) axi-planar ($p_{x}+i\beta_{p}%
p_{y})$ state with $\beta_{p}$ depending on the momentum $p$, which breaks
rotational symmetry completely. Here $U(1)$ stands for the rotation about the
axis, and $S^{2}$ stands for the orientation about the unit sphere. We note
that the axial phase has additional gauge-orbital symmetry
\cite{Jason,Volovik}, and therefore has coreless vortex structure as in $^{3}%
$He-A phase \cite{Jason_vortex}.
\begin{figure}[t]
\begin{center}
\includegraphics[width = 0.99 \columnwidth]{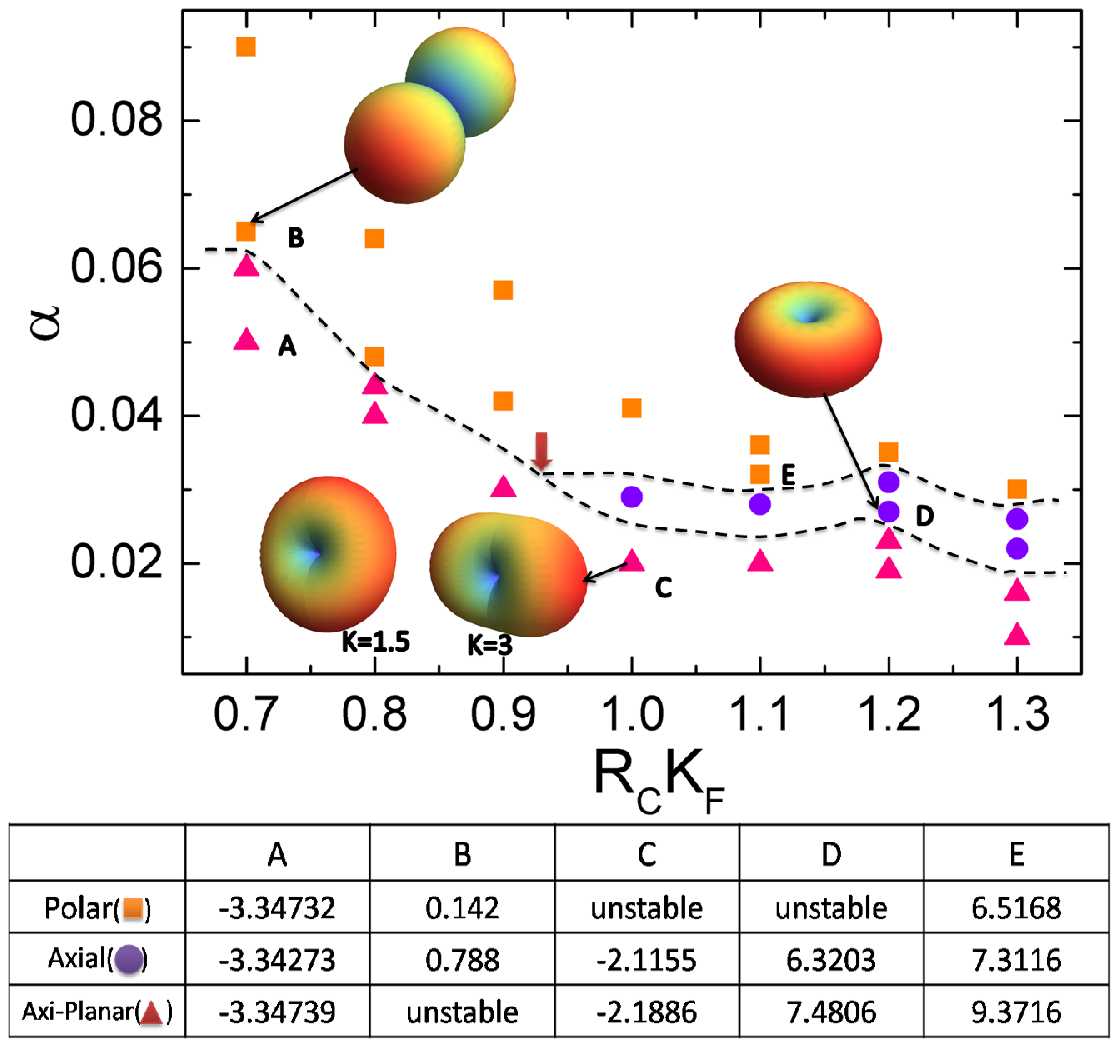}
\end{center}
\caption{(Color online) Quantum phase diagram for $p$-wave superfluid of
Rydberg-dressed fermions. Three different pairing phases (see the text) are
identified: polar ($\square$), axial ($\bigcirc$), and axi-planar
($\bigtriangleup$) states. Dashed lines are eye-guiding for the phase boundary
and insets show the 3D spherical plot of the modulus of the orientation
function, i.e. $|\Psi_{k}(\hat{k})|$. For the axi-planar state, we draw the
modulus for two different momenta, $k/k_{F}=1.5$ and $3$, as an example. The
table shows the corresponding total energies (in unit of Fermi energy) for
five representative points indicated in quantum phase diagram. Here,
"unstable" means such configuration is just a saddle point. The thick arrow
indicates the possible position of tri-critical point.}%
\label{fig3}%
\end{figure}

{\textit{Quantum phase diagram:}} Using the symmetry classification above, we
may represent the ground state as the following general configuration:
\begin{align}
\Psi_{k}(\hat{k})  &  =\cos\gamma_{k}Y_{1,1}(\hat{k})+\sin\gamma_{k}%
Y_{1,-1}(\hat{k})\nonumber\\
&  \propto\sin(\gamma_{k}-\pi/4)k_{x}-\cos(\gamma_{k}-\pi/4)ik_{y} \label{Psi}%
\end{align}
for $0\leq\gamma_{k}\leq\pi/4$. Note that all the phases mentioned above can
be expressed by properly rotating the coordinate axis (for example, polar
($p_{y}$) state can be obtained by using $\gamma_{k}=\pi/4$). From our full
numerical calculation (see Fig. \ref{fig2}(b) as an example), we find that the
momentum dependence of $\gamma_{k}$ is small, but finite ($\sim15\%$). In the
context of $^{3}$He system \cite{Anderson61,Leggett}, it was generally assumed
that $\Psi_{k}(\hat{k})$ is independent of the momentum magnitude, $k$,
because pairing occurs only near Fermi surface. However, as clearly shown
below, this assumption fails to explain the full phase diagram here, due to
the finite-range nature of the Rydberg-dressed interaction.

We calculate the ground state phase diagram via the following two steps: we
first solve the gap equation (Eqs. (\ref{reggap})) for a given density
self-consistently by iterating the initial configuration until convergence.
The gap symmetry is specified in the initial configuration for polar and axial
states. As for the axi-planar state, the relative strength of different
components (i.e. the value of $\gamma_{k}$) evolves self-consistently until
convergence. Next, we use the obtained gap function to calculate the
corresponding ground state energy (see Supplementary Material) for determining
the most favorite configuration. The finite temperature phases diagram can be
also obtained similarly by minimizing the total free energy.

In Fig. \ref{fig3} we show the obtained quantum phase diagram of the $p$-wave
superfluid, as a function of the soft-core radius, $R_{c}k_{F}$ and the
dimensionless interaction strength, $\alpha$, where $\alpha\equiv Mk_{F}%
^{-2}V_{0}/(2\pi)^{3}$ is the ratio of the interaction energy to the Fermi
energy. Some remarks can be addressed on the quantum phase diagram: (1) In the
parameter regime we studied, axi-planar state with a constant $\gamma_{k}%
\neq0,\pi/4$ is \textit{never} stabilized to be a ground state, as will be
explained in details below. (2) In the regime of weaker interaction (bottom
left), the axi-planar ($p_{x}+i\beta_{p}p_{y}$) seems dominant, while the
energy difference between various configurations is very small (see the
table). The momentum dependence of $\beta_{p}$ makes the gap function having
different symmetries at different $p$, as shown in the insets of Fig.
\ref{fig3}. (3) In the regime of stronger interaction (top right), both of
polar and axial phase can be found to be the true ground state. (4) In the
regime of even stronger interaction, however, our calculation cannot converge
efficiently and therefore cannot exclude the possibility of reentrant
transition, as mentioned in Refs. \cite{Leggett}.

To qualitatively understand the quantum phase diagram, it is better to start
from the Ginzburg-Landau (GL) free energy ($F_{\mathrm{GL}}$) with a small
order parameter just below $T_{c}$. For simplicity, we just show the leading
order expansion of the free energy for a constant $\gamma_{k}=\gamma$ (see
Supplementary Material):
\begin{align}
&  \frac{F_{GL}}{\mathcal{V}}=\int_{k}\tilde{f}_{2}(k)\Delta(k)^{2}%
-\int_{k_{1}}\int_{k_{2}}\tilde{g}_{2}(k_{1},k_{2})\Delta(k_{1})\Delta
(k_{2})\nonumber\\
&  +\left[  -\int_{k}\tilde{f}_{4}(k)\Delta(k)^{4}+2\int_{k_{1}}\int_{k_{2}%
}\tilde{g}_{4}(k_{1},k_{2})\Delta(k_{1})\Delta(k_{2})^{3}\right] \nonumber\\
&  \times\left(  2+\sin^{2}(2\gamma)\right)  +\mathcal{O}(\Delta^{6})
\label{F_GL3}%
\end{align}
where $\int_{k}\equiv\int_{0}^{\infty}\frac{k^{2}dk}{(2\pi)^{3}}$ and
functions, $\tilde{f}_{2}$, $\tilde{f}_{4}$, $\tilde{g}_{2}$ and $\tilde
{g}_{4}$ are defined explicitly in the Supplementary Material (we neglect the
temperature dependence for simplicity). We first find that the $T_{c}$, which
is determined solely by the sign of the quadratic order, is independent of the
gap symmetry (i.e., $\gamma$). Secondly, when $T<T_{c}$ for a finite
$\Delta(k)$, the quartic order is minimized either at $\gamma=0$ (axial state)
or $\gamma=\pi/4$ (polar state), depending on the sign of its coefficient.
This explains why the axi-planar state with constant $\gamma_{k}=\gamma$
cannot be favoured in the ground state calculation. The polar state observed
in strong interaction regime probably has to be understood by including higher
order effects.

In the weak interaction regime, on the other hand, we find an axial state
($\gamma=0$) can be energetically favoured, if only the momentum transfer
($\sim R_{c}^{-1}$) is smaller than Fermi momentum, $k_{F}$, so that pairing
occurs near Fermi surface, similar to the $^{3}$He-$A$ phase studied by
Anderson and Morel \cite{Anderson61}. This condition is satisfied in our
system when $R_{c}k_{F}$ is large (i.e. the bottom right corner of Fig.
\ref{fig3}). The phase transition between the polar and the axial state is
expected to be first order due to breaking time reversal symmetry of the axial state.

Finally, in the weak interaction and smaller $R_{c}$ regime, the momentum
transfer can be comparable to $k_{F}$ so that the possibility of momentum
mixing cannot be neglected in the free energy (see Eq. (\ref{F_GL3})). This
effect is expected to be important especially in the weak interaction limit,
because energies of various configurations are very close to each other. This
explains why axi-planar state has momentum dependence in $\gamma_{k}$ (i.e.
gap function at different relative momentum are mixed due to scattering).

\begin{figure}[t]
\begin{center}
\includegraphics[width = 0.99 \columnwidth]{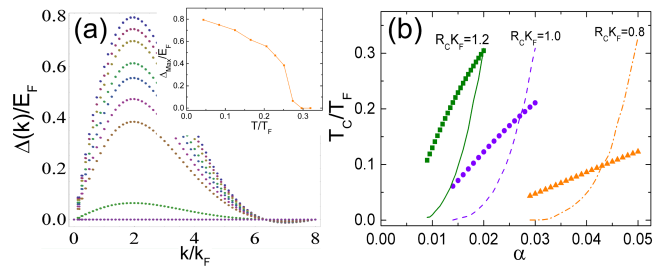}
\end{center}
\caption{(color online) (a) Calculated gap function, $\Delta(k)$, for a polar
state in various temperature. Here we choose $\alpha=0.03$ and $R_{c}k_{F}=1$,
and each of lines corresponds to the temperature from $T/E_{F}=0.044$ to $0.3$
with an interval $0.032$ from top to bottom. The inset shows how its maximum
value changes as function of temperature. (b) The critical temperature as
function of interaction strength, $\alpha$, for various $R_{c}k_{F}$. Lines
are the results by self-consistently solving the gap equation, while symbols
are the results obtained by Eq. (\ref{Tc1}) within the GM approach (see the
text).}%
\label{fig4}%
\end{figure}

{\textit{Transition temperature, $T_{c}$:}} We now estimate the critical
temperature of the $p$-wave superfluid. We first solve the BCS gap equation
(Eq. (\ref{reggap})) and the density equation self-consistently. The
calculated gap function $\Delta(k)$ of a polar state are shown in Fig.
\ref{fig4}(a) as a function of the momentum $k/k_{F}$ for various temperature
$T/E_{F}$. In the inset, we show how the maximum value of the gap function
decreases to zero when the temperature is increased from zero to above $T_{c}$.

In Fig. \ref{fig4}(b), we show the obtained $T_{c}$ as a function of the
interaction strength, $\alpha$, for three different values of the soft-core
radius. As expected in a standard BCS result, $T_{c}$ increases exponentially
as a function of $\alpha$. In the large $\alpha$ or $R_{c}$ regime, $T_{c}$
can be comparable to Fermi energy. Although the BCS theory may not be
justified in the strong interaction regime quantitatively, our results may
still provide useful information because our self-consistent calculation has
included the renormalization of the chemical potential, and the parameters
here are still within the weak interaction limit for the the first Born
approximation without two-body bound states.

Besides of the numerical results, we can also estimate $T_{c}$ through the
following analytic results by concentrating on the contribution of pairing
near Fermi surface (see Supplementary Material):
\begin{equation}
\frac{T_{c}}{E_{F}}=\frac{2\,e^{\gamma}\bar{\omega}}{\pi}\exp\left(  -\frac
{1}{2\nu_{0}V_{1}(k_{F},k_{F})}\right)  , \label{Tc1}%
\end{equation}
where $\nu_{0}$ is the density of state at Fermi surface. The energy scale,
$\bar{\omega}$, can be estimated within the Gor'kov and Melik-Barkhudarov (GM)
approach \cite{GM1961,BaranovPRA2002}, which has taken into account the
non-local correlation. Note that GM approach is important especially for
finite-range interaction here, because it involves the pairing scattering away
from the Fermi surface. The details of calculating $T_{c}$ within the GM
approach can be found in the Supplementary Material. In Fig. \ref{fig4}(b), we
show results of full numerical and GM methods together for comparison, and
find that the transition temperature of the later is enhanced in the weak
interaction limit, while it represents a strong suppression in the strong
interaction regime. Such discrepancy has been observed in earlier works of 2D
polar molecules \cite{SantosPRL2010}. Within the parameter regime we are
interested in the quantum phase diagram (say, $\alpha<0.04$ and $0.8<R_{c}%
k_{F}<1.2$), it seems very promising to have at least $T_{c}\sim0.05-0.1E_{F}%
$. We note that an even higher $T_{c}$ can be expected if one increases the
interaction strength ($\alpha$) or interaction range ($R_{c}$) in a realistic
experimental parameters (see below).

{\textit{Discussion:}} Finally, we discuss some experimental conditions for
preparing and observing of $p$-wave superfluid in Rydberg-dressed fermions.
For the experimental preparation, one may consider $^{6}$Li by excited to a
low-lying Rydberg state $n=30$, which has an average van der Waal coefficient,
$\bar{C}_{6}\sim2$ GHz-$\mu$m$^{6}$ \cite{SingerJPB2005}. For $\delta
_{\mathrm{eff}}=2\pi\times1$ GHz, and $\Omega_{\mathrm{eff}}=2\pi\times100$
MHz, we have soft-core radius, $R_{c}\sim0.74$ $\mu$m. If considering a low
density regime, $n\sim10^{12}$ cm$^{-3}$ (i.e. $E_{F}\sim12.7$ kHz and
$k_{F}\sim3.9$ $\mu$m$^{-1}$), we may have $\alpha=0.012$. Values of $\alpha$
and $R_{c}$ can be tuned by external fields in a wide range.

For a realistic Fermi gas in a harmonic trap, one may apply local density
approximation so that the pairing wavefunction in the trap center (higher
density) can be different from that near the edge. Near the boundary of a
chiral $p$-wave (i.e. $p_{x}+ip_{y}$) phase, it is then possible to find a
localized Majorana mode. More systematic calculation needs to be done by
self-consistently solving the full Bogoliubov-de Gennes equation, and will be
presented in other places.

To measure the $p$-wave superfluid we discuss here, one can first identify the
superfluidity by observing vortices through a rotating experiment below
$T_{c}$. The anisotropic $p$-wave pairing may not be easily measured from the
gap anisotropy, since the corresponding frequency is in the radio-wave regime,
where the angle resolution is very poor. But it may be observed from the
anisotropic noise correlation \cite{Ehud}, which should show different results
along the axial direction and the planar direction if the atomic cloud is
trapped in an anisotropic trap. The ground states of different pairing can be
also distinguished by their vortices structure, since the symmetries of their
order parameters are different. For example, axial phase is equivalent to
$^{3}$He-$A_{1}$ phase \cite{Volovik} or the "ferromagnetic" state of spin-1
condensate \cite{Jason}. As a result, it should have coreless vortices
\cite{Volovik}, which can be observed through interference pattern. Polar
phase is equivalent to $^{3}$He-$\beta$ phase, and therefore shall have
standard vortices, similar to $^{3}$He-$B$ phase. As for the vortices in the
axi-planar phase, it has no counterpart in other $p$-wave systems and needs
more careful classification. But its anisotropic feature should makes the
vortex structure highly depending on the geometry of the trapping potential.
We leave the detailed calculation of the vortex structure in the future.

{\textit{Summary:}} In summary, we systematically investigate the novel BCS
superfluid pairing of Rydberg-dressed identical fermions, identifying three
distinct ground states according to their symmetries. Compared to other
$p$-wave candidates in ultracold atoms/molecules, the finite-range interaction
here leads not only rich quantum phase diagram as shown in Fig. \ref{fig3},
but also enhances the transition temperature in $p$-wave channel. Our studies
therefore pave the way to study the exotic $p$-wave superfluid and related
phenomena in the future.

We acknowledge the helpful discussion with C. S. Adams, T.-L. Ho, A. Leggett,
C.-Y. Mou, T. Pohl, G. Shlyapnikov, W.-C. Wu and S. K. Yip. This works is
supported by NSC and NCTS in Taiwan.


\end{document}


\title{Supplementary material:\\$P$-wave Superfluidity by Blockade Effects in a Rydberg-Dressed Fermi Gas}
\email{stevenxiongbo@gmail.com}
\author{Bo Xiong}
\affiliation{Physics Department, National Tsing Hua University, Hsinchu, Taiwan}
\author{H. H. Jen}
\affiliation{Physics Department, National Tsing Hua University, Hsinchu, Taiwan}
\author{Daw-Wei Wang}
\affiliation{Physics Department, National Tsing Hua University, Hsinchu, Taiwan}
\affiliation{Physics Division, National Center for Theoretical Sciences, Hsinchu, Taiwan}
\affiliation{Frontier Research Center on Fundamental and Applied Sciences of Matter,
National Tsing Hua University, Hsinchu, Taiwan}

\maketitle

\subsection{A: Parameter regime without two-body bound state}

\begin{figure}[t]
\begin{center}
\includegraphics[width = 1.15\columnwidth]{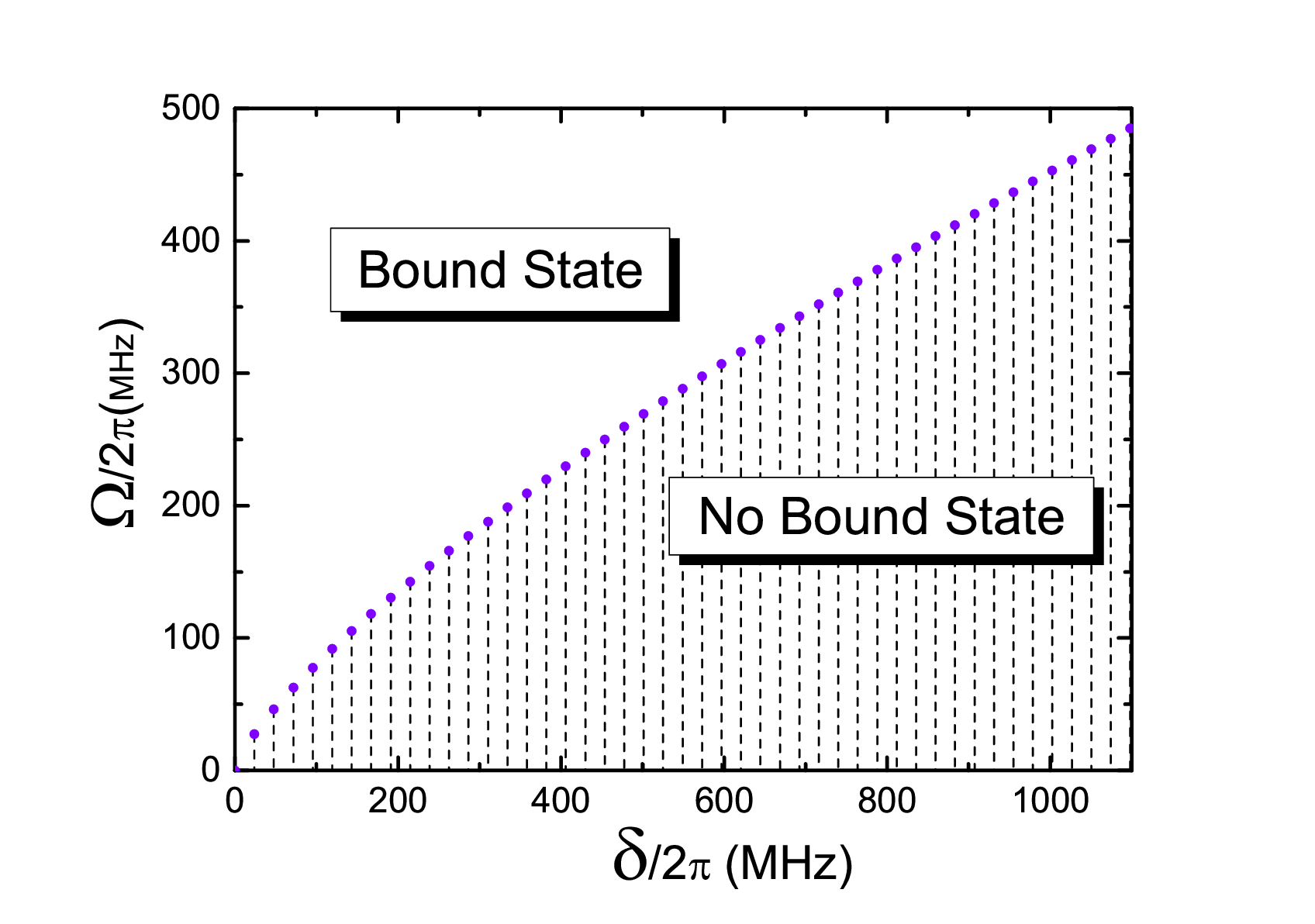}
\end{center}
\caption{(Color online) Parameter regions without a two-body bound state for a
Rydberg-dressed interaction, as a function of the effective Rabi frequency
$\Omega_{\mathrm{eff}}$ and laser detuning $\delta_{\mathrm{eff}}$. Results
are based on the parameter of $^{6}$Li atoms excited to a Rydberg state of
$n=30$.}%
\label{fig5}%
\end{figure}

In the system of Rydberg-dressed polarized Fermi gases, the effective two-body
interaction is given by an isotropic soft-core potential $V\left(  r\right)  $
(See the Eq. (1) in the main text). As a result, the two-body bound state, if
exists, can be described by the 3D Schr\"{o}dinger equation with the
separation of variables, i.e. $\psi(\mathbf{r})=r^{-1}\sum_{nlm}%
u_{nl}(r)Y_{lm}(\hat{r})$, where $\mathbf{r}\equiv\mathbf{r}_{1}%
-\mathbf{r}_{2}$ is the relative position between two particles, and
$u_{nl}(r)$ has to satisfy the following equation ($\hbar\equiv1$):
\begin{gather}
-\frac{1}{2M_{r}}\frac{d^{2}}{dr^{2}}\left[  u_{n,l}\left(  r\right)  \right]
+\left[  \frac{-V_{0}}{1+\left(  r/R_{c}\right)  ^{6}}+\frac{l\left(
l+1\right)  }{2M_{r}r^{2}}\right]  u_{n,l}\left(  r\right) \nonumber\\
=E_{n,l}u_{n,l}\left(  r\right)  . \label{b0}%
\end{gather}
Here $M_{r}=M/2$ is the reduced mass, and $E_{n,l}$ is the bound state
eigen-energy of principle quantum number $n$ and angular momentum quantum
number, $l$. For the purpose of this paper, we are looking at the situation
when the attractive interaction strength ($V_{0}>0$) is so weak that at least
a two-body bound state does not exist. Bound states with more than two
identical fermions, although in principle possible, but are very unlikely to
be relevant due to Pauli exclusion principle. The criteria for such condition
(no two-body bound states) can be easily obtained by taking the lowest energy,
$E_{1,1}=0$ (for identical fermionic atoms, the bound state must be
antisymmetric, i.e. $l=1,3,\cdots$), such that the above equation determining
the criteria of a bound state becomes ($\tilde{r}\equiv r/R_{c}$)
\begin{equation}
-\frac{d^{2}u_{n,l}\left(  \tilde{r}\right)  }{d\tilde{r}^{2}}+\left[
\frac{-\tilde{V}_{0}}{1+{\tilde{r}}^{6}}+\frac{2}{\tilde{r}^{2}}\right]
u_{n,l}\left(  \tilde{r}\right)  =0, \label{b3}%
\end{equation}
where
\begin{equation}
\tilde{V}_{0}\equiv2M_{r}R_{c}^{2}V_{0}. \label{b4}%
\end{equation}
is a dimensionless parameter. As a result, the functional relationship between
the effective Rabi frequency, $\Omega_{\mathrm{eff}}$, and effective detuning,
$\delta_{\mathrm{eff}}$, is:%
\begin{equation}
\Omega_{\mathrm{eff}}=\left(  \frac{\tilde{V}_{0}}{MC_{6}^{1/3}}\right)
^{1/4}(2\delta_{\mathrm{eff}})^{5/6},
\end{equation}
below which we should have no two-body bound state between identical fermions.
Numerical calculation gives $\tilde{V}_{0}=7.78.$


\subsection{B: Ginzburg-Landau equation for $p$-wave superfluid}

The total free energy of a $p$-wave superfluid with the BCS mean field
approximation has been well-known in the literature (see for example, Ref.
\cite{Leggett}), and is composed of kinetic energy, entropy of
quasi-particles, and interaction energy as following:
\begin{widetext}
\begin{eqnarray}
F&=&\sum_{\bfk}\left[ \xi_{\bfk}-E_\bfk +\frac{\left\vert\Delta_\bfk\right\vert^{2}}{E_\bfk}\tanh \frac{E_\bfk}{2T}-2T\ln\left( 1+e^{-E_\bfk/T} \right)\right]
+\frac{1}{{\cal V}}\sum_{\mathbf{k,k}^{\prime }}V_{\bfk-\bfk'}\left[ \frac{\Delta_\bfk}{2E_\bfk }\tanh \frac{E_\bfk}{2T}\right] \left[ \frac{\Delta^\ast_{\bfk'} }{2E_{\bfk'} }\tanh
\frac{E_{\bfk'}}{2T}\right],
\label{F_total}
\end{eqnarray}
where we keep the full expression of the gap function, $\Delta_{\mathbf{k}}$ as well as the interaction matrix element, $V_{\bfk-\bfk'}$.
The Ginzburg-Landau free energy ($F_{\mathrm{GL}}$) is then obtained by expanding the full free energy above to the leading orders of small $|\Delta_{\mathbf{k}}|$. After some straightforward algebra, we can obtain
$F_{\mathrm{GL}}$ to the quartic order:
\begin{eqnarray}
F_{\rm GL}\left(\left\{\Delta_\bfk,\Delta^\ast_\bfk\right\};T\right)/{\cal V}&=&
\frac{1}{{\cal V}}\sum_\bfk f_{2}(k;T)|\Delta_\bfk|^2
-\frac{1}{{\cal V}^2}\sum_{\bfk_1,\bfk_2}g_{2}(\bfk_1,\bfk_2;T)\Delta^\ast_{\bfk_1}\Delta_{\bfk_2}
-\frac{1}{{\cal V}}\sum_\bfk f_{4}(k;T)|\Delta_\bfk|^4
\nonumber\\
&&+\frac{1}{{\cal V}^2}\sum_{\bfk_1,\bfk_2}g_{4}(\bfk_1,\bfk_2;T)
\left[\Delta_{\bfk_1}\Delta_{\bfk_2}^\ast+\Delta_{\bfk_1}^\ast\Delta_{\bfk_1}\right]|\Delta_{\bfk_2}|^2
+{\cal O}\left(\Delta^6\right),
\label{F_GL0}
\end{eqnarray}
\end{widetext}
where we have defined the following functions:
\begin{align}
f_{2}(k;T)  &  \equiv\frac{\tanh(|\xi_{\mathbf{k}}|/2T)}{2|\xi_{\mathbf{k}}%
|}\\
f_{4}(k;T)  &  \equiv\frac{3}{8|\xi_{\mathbf{k}}|^{3}}\tanh\left(  \frac
{|\xi_{\mathbf{k}}|}{2T}\right) \nonumber\\
&  -\frac{3}{16T|\xi_{\mathbf{k}}|^{2}}\mathrm{sech}^{2}\left(  \frac
{|\xi_{\mathbf{k}}|}{2T}\right) \\
g_{2}(\mathbf{k}_{1},\mathbf{k}_{2};T)  &  \equiv-V_{\mathbf{k}_{1}%
-\mathbf{k}_{2}}f_{2}(k_{1};T)f_{2}(k_{2};T)\\
g_{4}(\mathbf{k}_{1},\mathbf{k}_{2};T)  &  \equiv-V_{\mathbf{k}_{1}%
-\mathbf{k}_{2}}f_{2}(k_{1};T){f}^{\prime}_{4}(k_{2};T)\\
{f}^{\prime}_{4}(k;T)  &  \equiv\frac{\mathrm{sech}^{2}\left(  \frac
{|\xi_{\mathbf{k}}|}{2T}\right)  \left[  T\mathrm{sinh}\left(  \frac
{|\xi_{\mathbf{k}}|}{T}\right)  -|\xi_{\mathbf{k}}| \right]  }{8T|\xi
_{\mathbf{k}}|^{2}}.
\end{align}
Note that all of these function are regular and well-defined near the Fermi
surface, $\xi_{\mathbf{k}}=\mathbf{k}^{2}/2M-\mu\rightarrow0$, and $f_{2}$,
$f_{4}$ and ${f}^{\prime}_{4}$ are all dependent on the magnitude of
$\mathbf{k}$ only. We have selected the sign properly, so that all the
functions, $f_{2}$, $f_{4}$, ${f}^{\prime}_{4}$, $g_{2}(\mathbf{k}%
_{1},\mathbf{k}_{2};T)$, and $g_{4}(\mathbf{k}_{1},\mathbf{k}_{2};T)$ are
positive defined near Fermi surface (note that $V_{\mathbf{k}-\mathbf{k}%
^{\prime}}$ is negative for an attractive interaction here).

Since only $p$-wave component is relevant in our present system, we can use
$V_{\mathbf{k}_{1}-\mathbf{k}_{2}}\approx-V_{1}(k,k^{\prime})\sum_{m=0,\pm
1}Y_{1,m}^{\ast}(\hat{k})Y_{1,m}(\hat{k}^{\prime})$ and consider the gap
configuration ((See the Eq. (3) in the main text)). Integrating out the
angular degree of freedom, we can rewrite Eq. (\ref{F_GL0}) as following:
\begin{widetext}
\begin{eqnarray}
&&F_{\rm GL}\left(\Delta(k),\{\alpha_m\};T\right)/{{\cal V}}
\nonumber\\
&&=\int_0^\infty\frac{k^2dk}{(2\pi)^3} \tilde{f}_{2}(k;T)\Delta(k)^2
-\int_0^\infty\frac{k_1^2dk_1}{(2\pi)^3}\int_0^\infty\frac{k_2^2dk_2}{(2\pi)^3}\tilde{g}_{2}(k_1,k_2;T)\Delta(k_1)\Delta(k_2)
\nonumber\\
&&-\int_0^\infty\frac{k^2dk}{(2\pi)^3}\tilde{f}_{4}(k;T)\Delta(k)^4
\sum_{m_1,\cdots,m_4}A_{m_1,m_2,m_3,m_4}
\beta^{m_1}_{k}{}^\ast\beta^{m_2}_k{}^\ast\beta^{m_3}_k\beta^{m_4}_k
\nonumber\\
&&+\int_0^\infty\frac{k_1^2dk_1}{(2\pi)^3}\int_0^\infty\frac{k_2^2dk_2}{(2\pi)^3}\left[\tilde{g}_{4}(k_1,k_2;T)\Delta(k_1)\Delta(k_2)^3 \sum_{m_1,\cdots,m_4}A_{m_2,m_3,m_4,m_1}
\beta^{m_1}_{k_1}\beta_{k_2}^{m_2}{}^\ast\beta_{k_2}^{m_3}{}^\ast\beta_{k_2}^{m_4}+c.c\right]+{\cal O}\left(\Delta^6\right),
\label{F_GL1}
\end{eqnarray}
\end{widetext}
where we have used the renormalization condition: $\sum_{m=0,\pm1}|\beta
_{k}^{m}|^{2}=1$, and
\begin{align}
\tilde{f}_{2}(k;T)  &  \equiv f_{2}(k;T)\\
\tilde{f}_{4}(k_{1},k_{2};T)  &  \equiv\frac{3}{20\pi}f_{4}(k;T)\\
\tilde{g}_{2}(k_{1},k_{2};T)  &  \equiv V_{1}(k_{1},k_{2})f_{2}(k_{1}%
;T)f_{2}(k_{2};T)\\
\tilde{g}_{4}({k}_{1},{k}_{2};T)  &  \equiv\frac{3}{20\pi}V_{1}(k_{1}%
,k_{2})f_{2}(k_{1};T){f}_{4}^{\prime}(k_{2};T)
\end{align}
\begin{align}
&  A_{m_{1},m_{2},m_{3},m_{4}}\nonumber\\
&  \equiv\frac{20\pi}{3}\int d\Omega_{k}Y_{1,m_{1}}^{\ast}(\hat{k})Y_{1,m_{2}%
}^{\ast}(\hat{k})Y_{1,m_{3}}(\hat{k})Y_{1,m_{4}}(\hat{k})\\
&  =2\mbox{ for }m_{1}+m_{2}=m_{3}+m_{4}=\pm2\nonumber\\
&  =1\mbox{ for }m_{1}+m_{2}=m_{3}+m_{4}=\pm1\nonumber\\
&  =3\mbox{ for }m_{1}=m_{2}=m_{3}=m_{4}=0\nonumber\\
&  =-1\mbox{ for }m_{1}+m_{2}=m_{3}+m_{4}=0,\mbox{ and }\sum_{i}%
|m_{i}|=2\nonumber\\
&  =2\mbox{ for }m_{1}+m_{2}=m_{3}+m_{4}=0,\mbox{ and }\sum_{i}|m_{i}%
|=4\nonumber\\
&  =0\mbox{ otherwise}
\end{align}

To understand the physical meaning of above GL free energy, we first note that
the quadratic orders depends only on the "amplitude" of the gap function, i.e.
$\Delta(k)$, and hence $T_{c}$ obtained by $\delta\mathcal{F}_{\mathrm{GL}%
}/\delta\Delta^{\star}(k)|_{\Delta(k)\rightarrow0}=0$ is independent of the
orientation. In order understand how the orientation of the order parameter
($\Delta_{\mathbf{k}}$) may affect the free energy, we take the orientation
configuration ((See the Eq. (4) in the main text)), so that
\begin{align}
&  \sum_{m_{1},\cdots,m_{4}}A_{m_{1},m_{2},m_{3},m_{4}}\beta_{k}^{m_{1}}%
{}\beta_{k}^{\ast m_{2}}{}\beta_{k}^{\ast m_{3}}\beta_{k}^{m_{4}}\nonumber\\
&  =2(\cos^{4}\gamma_{k}+\sin^{4}\gamma_{k})+8\cos^{2}\gamma_{k}\sin^{2}%
\gamma_{k}\nonumber\\
&  =2+\sin^{2}(2\gamma_{k})
\end{align}
and
\begin{align}
&  \sum_{m_{1},\cdots,m_{4}}A_{m_{2},m_{3},m_{4},m_{1}}^{m_{1}}{}\beta^{k_{1}%
}\beta_{k_{2}}^{m_{2}}{}^{\ast}\beta_{k_{2}}^{m_{3}}{}^{\ast}\beta_{k_{2}%
}^{m_{4}}\nonumber\\
&  =2(\cos\gamma_{k_{1}}\cos^{3}\gamma_{k_{2}}+\sin\gamma_{k_{1}}\sin
^{3}\gamma_{k_{2}})\nonumber\\
&  +4\sin(\gamma_{k_{1}}+\gamma_{k_{2}})\sin\gamma_{k_{2}}\cos\gamma_{k_{2}}%
\end{align}

We note that, if we consider a simple case when the amplitude of the gap
function is completely decoupled from it orientation (i.e. $\gamma_{k}=\gamma$
is independent of momentum magnitude, $k$, or equivalently, $\beta_{k}%
^{m}=\beta^{m}$ for all $k$), we can simplify above results and obtain
\begin{align}
&  \sum_{m_{1},\cdots,m_{4}}A_{m_{1},m_{2},m_{3},m_{4}}\beta_{k}^{m_{1}}%
{}^{\ast}\beta_{k}^{m_{2}}{}^{\ast}\beta_{k}^{m_{3}}\beta_{k}^{m_{4}%
}\nonumber\\
&  =\sum_{m_{1},\cdots,m_{4}}A_{m_{2},m_{3},m_{4},m_{1}}\beta_{k_{1}}^{m_{1}%
}{}\beta_{k_{2}}^{m_{2}}{}^{\ast}\beta_{k_{2}}^{m_{3}}{}^{\ast}\beta_{k_{2}%
}^{m_{4}}\nonumber\\
&  =2+\sin^{2}(2\gamma).
\end{align}
The Ginzburg-Landau free energy for such ansatz then becomes a simple function
of $\gamma$:
\begin{widetext}
\begin{eqnarray}
&&F_{\rm GL}\left(\Delta(k),\gamma;T\right)/{{\cal V}}
\nonumber\\
&&=\int_0^\infty\frac{k^2dk}{(2\pi)^3} \tilde{f}_{2}(k;T)\Delta(k)^2
-\int_0^\infty\frac{k_1^2dk_1}{(2\pi)^3}\int_0^\infty\frac{k_2^2dk_2}{(2\pi)^3}\tilde{g}_{2}(k_1,k_2;T)\Delta(k_1)\Delta(k_2)
\nonumber\\
&&+\left[-\int_0^\infty\frac{k^2dk}{(2\pi)^3}\tilde{f}_{4}(k;T)\Delta(k)^4
+2\int_0^\infty\frac{k_1^2dk_1}{(2\pi)^3}\int_0^\infty\frac{k_2^2dk_2}{(2\pi)^3}\tilde{g}_{4}(k_1,k_2;T)
\Delta(k_1)\Delta(k_2)^3\right]\left(2+\sin^2(2\gamma)\right)+{\cal O}\left(\Delta^6\right)
\label{F_GL2}
\end{eqnarray}
\end{widetext}


\subsection{C: Transition temperature of $p$-wave superfluid within the BCS
theory}

In this section, we derive semi-analytic results for the transition
temperature ($T_{c}$) of $p$-wave pairing superfluid within the BCS theory. By
de-composting the interaction matrix element and the gap function into
different orbital angular momentum channels: $V_{\mathbf{k}-\mathbf{k}%
^{\prime}}=-\sum_{lm}^{\prime}{V}_{l}(k,k^{\prime})Y_{l,m}^{\ast}(\hat
{k}^{\prime})Y_{l,m}(\hat{k})$ and $\Delta_{\mathbf{k}}=\sum_{lm}^{\prime
}\Delta_{l,m}(k)Y_{l,m}(\hat{k})$, where $\sum_{lm}^{\prime}$ is summation
over all odd values of $l$ and $m=-l,\cdots,l$, we find that the major
contribution is from $p$-wave channel ($l=1$) as shown in Fig \ref{fig2}(a).
Since near $T_{c}$, pairing function of different channels are degenerate,
below we will assume a polar state, $\Delta_{\mathbf{k}}=\Delta_{1,0}%
Y_{1,0}(\hat{k})$, for the study of $T_{c}$. As a result, we may apply the
theory developed in Ref. \cite{BaranovPRA2002} in the context of fermionic
dipolar gas to investigate $T_{c}$, where the gap function is also a $p_{z}$
type polar state.

We start from expanding the right hand side gap equation to the leading order,
and rescale all quantities to dimensionless expression in unit of Fermi
wavevector ($k_{F}$) and Fermi energy ($E_{F}$), i.e. $\tilde{\mathbf{k}%
}\equiv\mathbf{k}/k_{F}$, $\tilde{k}\equiv k/f_{F}$, $\tilde{T}\equiv T/E_{F}%
$, $\tilde{\Delta}_{1}\equiv\Delta_{1,0}/E_{F}$, $\tilde{\nu}\equiv\mu/E_{F}$,
and $\alpha\tilde{V}_{1}(\tilde{k},\tilde{k}^{\prime})\equiv V_{1}%
(k,k^{\prime})k_{F}^{3}/2E_{F}/(2\pi)^{3}$, so that after integration of
angular part, gap equation reduces to
\begin{equation}
\tilde{\Delta}_{1}(\tilde{k})=\alpha\int_{0}^{\infty}\tilde{k}^{\prime
2}d\tilde{k}^{\prime}\frac{\tilde{V}_{1}(\tilde{k},\tilde{k}^{\prime}%
)}{|\tilde{k}^{\prime2}-\tilde{\mu}|}\tanh\left(  \frac{|\tilde{k}^{\prime
2}-\tilde{\mu}|}{2\tilde{T}}\right)  \tilde{\Delta}_{1}(\tilde{k}^{\prime})
\label{7}%
\end{equation}

In order to single out the main contribution in the rhs of Eq. (\ref{7}), we
introduce a characteristic frequency or energy $\bar{\omega}$ that obeys the
constraint $T\ll\bar{\omega}$ and is of the order of the Fermi energy. By
introducing reduced kinetic energy, $\tilde{\xi}\equiv\tilde{k}^{2}$, and the
density of state, $\nu(\tilde{\xi}) =\tilde{k}^{2}d\tilde{k}/d\tilde{\xi}$, we
thus rewrite the r.h.s of Eq. (\ref{7}) to be
\begin{widetext}
\begin{gather}
\tilde{\Delta}_{1}(\tilde{\xi})=\alpha\left(  \int\limits_{0}^{\tilde{\mu}-\bar{\omega}}
+\int\limits_{\tilde{\mu}-\bar{\omega}}^{\tilde{\mu}+\bar{\omega}}
+\int\limits_{\tilde{\mu}+\bar{\omega}}^{\infty}\right)  d\tilde{\xi}^{\prime}\nu(  \tilde{\xi}^{\prime})
\frac{\tilde{V}_{1}(\tilde{\xi},\tilde{\xi}^{\prime})}{\vert \tilde{\xi}'-\tilde{\mu}\vert }
\tanh\left(  \frac{\vert \tilde{\xi}'-\tilde{\mu}\vert}{2\tilde{T}}\right)
\tilde{\Delta}_{1}(\tilde{\xi}^{\prime})
\nonumber\\
=\alpha\left(  \int\limits_{-\tilde{\mu}}^{-\bar{\omega}}
+\int\limits_{-\bar{\omega}}^{+\bar{\omega}}
+\int\limits_{+\bar{\omega}}^{\infty}\right)  d\tilde{\xi}^{\prime}\tilde{\nu}(  \tilde{\xi}^{\prime})
\frac{\tilde{V}_{1}(\tilde{\xi},\tilde{\xi}^{\prime})}{\vert \tilde{\xi}'\vert }
\tanh\left(  \frac{\vert \tilde{\xi}'\vert}{2\tilde{T}}\right)
\tilde{\Delta}_{1}(\tilde{\xi}^{\prime}),
\label{8}
\end{gather}
\end{widetext}
where we have move the quasi-particle energy relative to the chemical
potential (i.e. when $\tilde{\xi}=0$), redefine $\tilde{\Delta}_{1}(\tilde
{\xi})=\tilde{\Delta}_{1}(\tilde{\xi}+\tilde{\mu})$, $\tilde{V}_{1}(\tilde
{\xi},\tilde{\xi}^{\prime})\equiv\tilde{V}_{1}(\tilde{\xi}+\tilde{\mu}%
,\tilde{\xi}^{\prime}+\tilde{\mu})$, and let $\tilde{\nu}(\tilde{\xi}%
)\equiv\nu(\tilde{\xi}+\tilde{\mu})=\frac{1}{2}\sqrt{\tilde{\xi}+\tilde{\mu}}$.

For the three integral range above, it is easy to see that the major
contribution comes from the quasi-particle excitation near the Fermi surface.
Following the standard treatment, we first introduce the following function:
\begin{equation}
G(\tilde{\xi}\mathbf{,}\tilde{\xi}^{\prime}) =\tilde{\nu}(\tilde{\xi}^{\prime
}) \tilde{V}_{1}(\tilde{\xi}+\tilde{\mu},\tilde{\xi}^{\prime}+\tilde{\mu
})\tilde{\Delta}_{1}(\tilde{\xi}^{\prime}), \label{G_F}%
\end{equation}
so that three integrals becomes:
\begin{align}
&  \alpha\int\limits_{-\bar{\omega}}^{+\bar{\omega}} d\tilde{\xi}^{\prime}
\frac{\tanh(\vert\tilde{\xi}^{\prime}\vert/2\tilde{T})}{\vert\tilde{\xi
}^{\prime}\vert} G(\tilde{\xi},0)\nonumber\\
&  +\alpha\left(  \int\limits_{-\tilde{\mu}}^{-\bar{\omega}} +\int
\limits_{+\bar{\omega}}^{\infty}\right)  d\tilde{\xi}^{\prime} \frac
{\tanh(\vert\tilde{\xi}^{\prime}\vert/2\tilde{T})}{\vert\tilde{\xi}^{\prime
}\vert}G(\tilde{\xi}\mathbf{,}\tilde{\xi}^{\prime})\nonumber\\
&  +\alpha\int\limits_{-\bar{\omega}}^{+\bar{\omega}} d\tilde{\xi}^{\prime}
\frac{\tanh(\vert\tilde{\xi}^{\prime}\vert/2\tilde{T})}{\vert\tilde{\xi
}^{\prime}\vert} \left[  G(\tilde{\xi}\mathbf{,}\tilde{\xi}^{\prime}%
)-G(\tilde{\xi},0)\right]  . \label{8_2}%
\end{align}
The last two lines are considered as a minor contribution, because they
involve the integration of particles away from Fermi surface. We will discuss
this part in Appendix D later.

The first line is the major contribution of the right hand side, and therefore
the transition temperature should be determined by the following equation:
\begin{align}
\tilde{\Delta}_{1}(\tilde{\xi})  &  =\alpha G(\tilde{\xi},0)\int
\limits_{-\bar{\omega}}^{\bar{\omega}}d\tilde{\xi}^{\prime}\frac{\tanh\left(
|\tilde{\xi}^{\prime}|/2\tilde{T}\right)  }{|\tilde{\xi}^{\prime}|}\nonumber\\
&  \simeq2\alpha\tilde{\nu}(0)\tilde{V}_{1}(\tilde{\xi},0)\tilde{\Delta}%
_{1}(0)\ln\left(  \frac{2\bar{\omega}e^{\gamma}}{\pi\tilde{T}}\right)  ,
\label{9}%
\end{align}
where we have evaluated the integral following the standard treatment (see
Refs. \cite{BaranovPRA2002,Fetter,GM1961}), where $\gamma=0.5772$ is the Euler
constant. Taking its result at Fermi surface, $\tilde{\xi}=0$, we can rid of
the gap function and obtain the following expression for the transition
temperature in the standard BCS theory:
\begin{equation}
\tilde{T}_{c}=\frac{2\,e^{\gamma}}{\pi}\bar{\omega}\exp\left(  -\frac
{1}{2\alpha\tilde{\nu}(0)\tilde{V}_{1}(0,0)}\right)  . \label{Tc2}%
\end{equation}

Note that the energy scale, $\tilde{\omega}$, including the contribution of
interaction away from the Fermi surface, has to be determined from the "minor"
part of Eq. (\ref{8_2}). When considering the momentum (or energy) dependence
of the gap function, Eq. (\ref{9}) can be applied to determine the behaviour
of $\tilde{\Delta}_{1}(\tilde{\xi})$ from the interaction, i.e.
\begin{equation}
\frac{\tilde{\Delta}_{1}(\tilde{\xi})}{\tilde{\Delta}_{1}(0)}=\frac{\tilde
{V}_{1}(\tilde{\xi},0)}{\tilde{V}_{1}(0,0)} \label{delta_V}%
\end{equation}
In our full numerical calculation, we can see that the gap function is indeed
proportional to the interaction matrix element as clearly shown in Fig. 2. The
close relationship between the gap function and the interaction matrix element
will help our further study in the determination of transition temperature.

\subsection{D: Calculation of the prefactor $\bar{\omega}$ beyond the BCS
theory}

In order to estimate the transition temperature more precisely, here we have
to determine the energy scale, $\bar{\omega}$, used in Eq. (\ref{8_2}) for the
separation of "major" contribution and "minor" contribution. Following the
seminar work by Gor'kov-Melik-Barkhudarov \cite{GM1961} and later extended
version for dipolar gas \cite{BaranovPRA2002}, we can estimate its
contribution first by replacing $\tanh(|\tilde{\xi}^{\prime}|/2\tilde{T})$
there via the step function and omitting the contribution from a narrow
interval $\left\vert \xi^{\prime}\right\vert \lesssim2\tilde{T}\ll\bar{\omega
}$. As a result, this minor contribution (the second and the third part of Eq.
(\ref{8_2})) becomes
\begin{widetext}
\begin{eqnarray}
&&\alpha\left(  \int\limits_{-\tilde{\mu}}^{-\bar{\omega}}
+\int\limits_{+\bar{\omega}}^{\infty}\right)  d\tilde{\xi}^{\prime} \frac{{\rm H}(\vert \tilde{\xi}'\vert/2\tilde{T})}{\vert\tilde{\xi}'\vert}G(\tilde{\xi},\tilde{\xi}^{\prime})
+\alpha\int\limits_{-\bar{\omega}}^{+\bar{\omega}}
d\tilde{\xi}^{\prime} \frac{{\rm H}(\vert \tilde{\xi}'\vert/2\tilde{T})}{\vert\tilde{\xi}'\vert}
\left[G(\tilde{\xi},\tilde{\xi}^{\prime})-G(\tilde{\xi},0)\right]
\nonumber\\
&=&\alpha\left(  \int\limits_{-\tilde{\mu}}^{+\bar{\omega}}
+\int\limits_{+\bar{\omega}}^{\infty}\right)  d\tilde{\xi}^{\prime} \frac{G(\tilde{\xi},\tilde{\xi}^{\prime})}{\vert\tilde{\xi}'\vert}
+\alpha\int\limits_{-\bar{\omega}}^{+\bar{\omega}}
d\tilde{\xi}^{\prime} \frac{1}{\vert\tilde{\xi}'\vert}
\left[G(\tilde{\xi},\tilde{\xi}^{\prime})-G(\tilde{\xi},0)\right]
=-\alpha\int\limits_{-\tilde{\mu}}^{\infty}d\tilde{\xi}^{\prime} \ln\left(\frac{|\tilde{\xi}'|}{\bar{\omega}}\right)\frac{dG(\tilde{\xi},\tilde{\xi}^{\prime})}{d|\tilde{\xi}'|},
\label{10_1}%
\end{eqnarray}
\end{widetext}
where the last form is obtained by the integration by parts for the three
integrals independently and then combining their results. This is important to
note that the integration of $\int_{-\bar{\omega}}^{+\bar{\omega}}$ is
composed by two divergent terms and therefore has to be treated differently.

Since the $T_{c}$ of Eq. (\ref{Tc2}) is based on the contribution of the
"major" part (the first line of Eq. (\ref{8_2})), it is therefore meaningful
to determine the unknown prefactor, $\bar{\omega}$, by requiring the
contribution from the "minor" part (the second and third lines of Eq.
(\ref{8_2}), or Eq. (\ref{10_1})) to be zero. Similar to the previous
treatment, when considering interaction of finite range, we should multiply
the density of state, $\tilde{\nu}(\tilde{\xi})$, and integrate these "minor"
part over all relevant energy. As a result, the equation to determine
$\bar{\omega}$ is given by the following form:
\begin{equation}
\int\limits_{-\tilde{\mu}}^{\infty}d\tilde{\xi}\int\limits_{-\tilde{\mu}
}^{\infty}d\tilde{\xi}^{\prime}\ln\left(  \frac{\vert\tilde{\xi}^{\prime}%
\vert}{\bar{\omega}}\right)  \frac{d}{d\vert\tilde{\xi}^{\prime}\vert}%
\tilde{G}( \tilde{\xi},\tilde{\xi}^{\prime}) =0, \label{14}%
\end{equation}
where we define
\begin{equation}
\tilde{G}( \tilde{\xi},\tilde{\xi}^{\prime})\equiv\tilde{\nu}(\tilde{\xi
})\tilde{\nu}( \tilde{\xi}^{\prime}) \tilde{V}_{1}(\tilde{\xi},\tilde{\xi
}^{\prime})\tilde{\Delta}_{1}(\tilde{\xi}^{\prime}). \label{14_0}%
\end{equation}

As a result, we can separate the term of $\bar{\omega}$ from others in Eq.
(\ref{14}) and obtain
\begin{align}
\ln\bar{\omega}  &  = \int\limits_{-\tilde{\mu}}^{\infty}d\tilde{\xi}%
\int\limits_{-\tilde{\mu} }^{\infty}d\tilde{\xi}^{\prime}\ln\vert\tilde{\xi
}^{\prime}\vert\frac{d}{d\vert\tilde{\xi}^{\prime}\vert}\tilde{G}( \tilde{\xi
},\tilde{\xi}^{\prime})\nonumber\\
&  \times\left[  \int\limits_{-\tilde{\mu}}^{\infty}d\tilde{\xi}%
\int\limits_{-\tilde{\mu}}^{\infty}d\tilde{\xi}^{\prime}\frac{d}{d\vert
\tilde{\xi}^{\prime}\vert}\tilde{G}( \tilde{\xi},\tilde{\xi}^{\prime})\right]
^{-1}\nonumber\\
&  =\frac{\int\limits_{-\tilde{\mu}}^{\infty}d\tilde{\xi}\int\limits_{-\tilde
{\mu} }^{\infty}d\tilde{\xi}^{\prime}\ln\vert\tilde{\xi}^{\prime}\vert\frac
{d}{d\vert\tilde{\xi}^{\prime}\vert}\tilde{G}( \tilde{\xi},\tilde{\xi}%
^{\prime})} {-2\int_{-\tilde{\mu}}^{\infty}\tilde{G}(\tilde{\xi},0)d\tilde
{\xi}}, \label{14_1}%
\end{align}
where we have evaluated the following integral directly
\begin{align}
&  \left(  \int\limits_{-\tilde{\mu}}^{0}+\int\limits_{0}^{\infty}\right)
d\tilde{\xi}^{\prime}\frac{d}{d\vert\tilde{\xi}^{\prime}\vert}\tilde{G}%
(\tilde{\xi},\tilde{\xi}^{\prime})\nonumber\\
&  =\int\limits_{0}^{\tilde{\mu}}d\tilde{\xi}^{\prime}\frac{d}{d\tilde{\xi
}^{\prime}}\tilde{G}( \tilde{\xi},-\tilde{\xi}^{\prime})+\int\limits_{0}%
^{\infty}d\tilde{\xi}^{\prime}\frac{d}{d\tilde{\xi}^{\prime}}\tilde{G}%
(\tilde{\xi},\tilde{\xi}^{\prime})\nonumber\\
&  =-2\tilde{G}(\tilde{\xi},0), \label{14_2}%
\end{align}

Replacing the definition of the function, $\tilde{G}$, and using Eq.
(\ref{delta_V}), we can rewrite Eq. (\ref{14_1}) to see how the energy scale,
$\bar{\omega}$, depends on the interaction matrix element and the gap function
explicitly:
\begin{align}
\ln\bar{\omega}  &  =\frac{\int\limits_{-\tilde{\mu}}^{\infty}d\tilde{\xi
}\,\tilde{\nu}(\tilde{\xi})\int\limits_{-\tilde{\mu}}^{\infty}d\tilde{\xi
}^{\prime}\ln|\tilde{\xi}^{\prime}|\frac{d}{d|\tilde{\xi}^{\prime}|}\left[
\tilde{\nu}(\tilde{\xi}^{\prime})\tilde{V}_{1}(\tilde{\xi},\tilde{\xi}%
^{\prime})\tilde{\Delta}_{1}(\tilde{\xi}^{\prime})\right]  }{-2\tilde{\nu
}(0)\tilde{\Delta}_{1}(0)\int_{-\tilde{\mu}}^{\infty}d\tilde{\xi}\,\tilde{\nu
}(\tilde{\xi})\tilde{V}_{1}(\tilde{\xi},0)}\nonumber\label{14_3}\\
&  =\frac{\int\limits_{-\tilde{\mu}}^{\infty}d\tilde{\xi}\,\tilde{\nu}%
(\tilde{\xi})\int\limits_{-\tilde{\mu}}^{\infty}d\tilde{\xi}^{\prime}%
\ln|\tilde{\xi}^{\prime}|\frac{d}{d|\tilde{\xi}^{\prime}|}\left[  \tilde{\nu
}(\tilde{\xi}^{\prime})\tilde{V}_{1}(\tilde{\xi},\tilde{\xi}^{\prime}%
)\tilde{V}_{1}(\tilde{\xi}^{\prime},0)\right]  }{-2\tilde{\nu}(0)\tilde{V}%
_{1}(0,0)\int_{-\tilde{\mu}}^{\infty}d\tilde{\xi}\,\tilde{\nu}(\tilde{\xi
})\tilde{V}_{1}(\tilde{\xi},0)},\nonumber\\
&
\end{align}

It is important to note that although the value of $\bar{\omega}$ looks like
depending on the second order of the interaction, but actually it is
independent on the "magnitude" of the interaction, i.e. $\alpha$, as clearly
seems from the original equation of such energy scale in Eq. (\ref{14}).
Therefore, calculating $\bar{\omega}$ within the GM approach does \textit{not}
mean that we have to include the second order Born approximation for the BCS
theory, which can still safely be assumed to be small. However, due to the
integral over all energy regime, it is very clear that the shape or the
sofe-core radius of the potential, $\tilde{V}_{1}$, about the Fermi energy can
be very crucial. In our present system, we find that $\bar{\omega}\sim0.55$
when $R_{c}k_{F}=1$.
